\documentclass[fleqn,usenatbib]{mnras}

\usepackage{mathptmx}

\usepackage[T1]{fontenc}
\usepackage{ae,aecompl}
\usepackage{comment}
\usepackage{verbatim}

\usepackage{graphicx}	
\usepackage{amsmath}	
\usepackage{amssymb}	
\usepackage{natbib}
\usepackage{bm}			
\RequirePackage{esint, amssymb, fancybox, empheq}

\newcommand{\nab}{\mbox{\boldmath $\nabla$}}

\title[Circularization of late--type binaries]{The circularization timescales of late--type binary stars}

\author[Caroline Terquem and Scott Martin]{
Caroline Terquem$^{1,2,3}$\thanks{caroline.terquem@physics.ox.ac.uk} and Scott Martin$^{2}$\thanks{scott.martin@univ.ox.ac.uk}
\\
Department of Physics, Oxford University, Keble Road, Oxford OX1 3RH, UK\\
University College, Oxford OX1 4BH, UK\\
Institut d'Astrophysique de Paris, Sorbonne Universit\'e, CNRS,
  UMR 7095, 98 bis boulevard Arago, F-75014, Paris, France \\
}

\date{Accepted XXX. Received YYY; in original form ZZZ}

\pubyear{2020}

\begin{document}
\label{firstpage}
\pagerange{\pageref{firstpage}--\pageref{lastpage}}
\maketitle

\begin{abstract}
We examine the consequences of, and apply, the formalism developed in Terquem (2021) for calculating the rate $D_R$ at which energy is exchanged between fast tides and convection.  In this previous work,  $D_R$  (which is proportional to the gradient of the convective velocity) was assumed to be positive in order to dissipate the tidal energy.  Here we argue that, even if energy is intermittently transferred from convection to the tides, it must ultimately return to the convective flow and transported efficiently to the stellar surface on the convective timescale.  This is consistent with, but much less restrictive than, enforcing $D_R>0$.     Our principle result is a calculation of the circularization timescale of late--type binaries, taking into account the full time evolution of the stellar structure.   We find that circularization is very efficient during the PMS phase, inefficient during the MS, and once again efficient when the star approaches the RGB. 
These results are in much better agreement with observations than earlier theories.  We also apply our 
formalism to hot Jupiters, and find that tidal dissipation in a Jupiter mass planet yields a circularization timescale of 1~Gyr for an orbital period of 3~d, also in good overall agreement with observations.  The approach here is novel, and the apparent success of the theory in resolving longstanding timescale puzzles is compelling.   
\end{abstract}



\section{Introduction}
\label{sec:intro}

Dissipation of tidal oscillations in the convective interior of late--type stars and giant planets is a key ingredient in determining  the orbital evolution of stellar binaries and that of the moons of giant planets.   Starting with \citet{Zahn1966}, numerous studies in the last 50~years have attempted to quantify the amount of energy  that can be extracted from the tides, assuming that convection acts as a turbulent viscosity \citep[and references therein]{Ogilvie2014}.    In this description, the rate of energy transfer between the tides and the convective flow is given by the coupling between the Reynolds stress associated with the convective velocities and the tidal shear flow, taking into account a reduction of dissipation when the convective turnover timescale $t_{\rm conv}$ is large compared to the tidal period $P$.   Tidal dissipation calculated this way is orders of magnitude too small to account for either the circularization period of late--type binaries, or the tidal dissipation factor of Jupiter and Saturn  inferred from the orbital motion of their  satellites.   

\citet[thereafter referred to as paper~I]{Terquem2021} revisited the interaction between tides and convection in the regime $P/t_{\rm conv} <1$.   From the form of the energy conservation equations, it was shown that traditional roles are actually reversed, with  the tides being the fluctuations and  convection being the mean flow.   This is because only the timescales on which the flows vary are relevant for  identifying the fluctuations and the mean flow.   This leads to
 the Reynolds stress being given by the correlation between the components of the tidal velocity, not that of the convective velocity.  The rate  $D_R$ at which energy is exchanged between the tides and the convective flow  is then determined  by the  coupling of this stress to the mean shear associated with the convective velocity.  In the regime $P/t_{\rm conv} >1$, the analysis presented in paper~I still applies but, in that case, the fluctuations are associated with the convective flow, and the tides are the mean flow.  This is the standard result of  \citet{Zahn1966}, which is very successful at reproducing the circularization timescales  of wide binaries containing giant stars,  which are fully convective and for which $P/t_{\rm conv} >1$ \citep{Verbunt1995}.   The results of paper~I  are therefore a step towards giving a unified description of the  interaction between tides and convection, which is likely to explain tidal dissipation in convective envelopes whether $P$ is larger or smaller than $t_{\rm conv} $.

The circularization of stellar binaries and the orbital evolution of the moons of giant planets are evidence that tidal energy is dissipated.  On this basis, it was assumed 
in paper~I that $D_R >0$.   The  tidal dissipation $Q$--factor  was calculated for Jupiter under this assumption and found to be in good agreement with recent observations.  Good agreement was also obtained for Saturn when using  models with a smaller mixing length parameter  than that adopted in stellar interiors.  

The eccentricity damping timescale, $t_e = -e \left( {\rm d} e / {\rm d} t \right)^{-1}$,  where $e$ is the eccentricity,  was also calculated  for a model of the current Sun  and for various orbital periods $P_{\rm orb}$.  This timescale was found to be about $40$~times too large to account for the observed circularization periods of late--type binaries.    However, $t_e$ is  {\em not} the circularization timescale for a given orbital period.     The time  $t_{\rm circ}$ it takes to circularize an orbit is the time up to which  $t^{-1}_e$ has to be integrated, starting at some initial value $t_0$, for the eccentricity to decrease from its  initial value by, say, an order of magnitude.    The calculation of $t_{\rm circ}$ for stellar binaries and hot Jupiters, using the new formalism presented in paper~I, is the object of the present paper.  

In section~\ref{sec:tcirc}, we review the results established in paper~I and explain how the circularization timescale is calculated.   We summarize the findings related to the exchange of energy between the tides and the convective flow in 
section~\ref{sec:energydissip}.  We also present an argument  in support of the idea that  energy is irreversibly transferred from the tides to the convective motions, which justifies using $D_R>0$.  
 In section~\ref{sec:evoltime}, we recall  the expression of the eccentricity damping timescale.    We then derive an expression for the circularization timescale in 
section~\ref{sec:circtime}.  We apply our results to late--type binaries in section~\ref{sec:sun}.  In 
section~\ref{sec:suncirc},  we calculate $t_{\rm circ}$ for binaries comprising two 1~M$_{\odot}$ stars, using MESA \citep{Paxton2011, Paxton2013,Paxton2015, Paxton2016, Paxton2018, Paxton2019} to generate models from an age of $\sim$ 0.1~Myr to an age of 12~Gyr.    In agreement with previous studies, circularization is found to  be very efficient during the pre--main sequence (PMS), and very inefficient during the main sequence (MS).  However, it becomes efficient again when the stars approach the red giant branch (RGB).   To illuminate these results, we describe the structure of the stars and their evolution in section~\ref{sec:convectionevol}.  We compare our results to observations in section~\ref{sec:compsun}.  We fix the time $t_0$ at which the integration for 
calculating $t_{\rm circ}$ starts by matching our results to the observed circularization period of PMS binaries.  We  find that our timescales are in broad agreement with observations for older clusters.
We note, however,  that the circularization periods $P_{\rm circ}$ determined from observed eccentricity-period distributions are very approximate, because they 
have been evaluated using a theory which is only valid for $e \ll 1$.  
In section~\ref{sec:hotjupiters}, we apply our results to binaries containing a solar--type star and a hot Jupiter.  We review observations in section~\ref{sec:jupitersobs}  and calculate circularization timescales in 
section~\ref{sec:tcircjupiters}, using MESA to produce models of a young Jupiter.  We find that tidal circularization is only efficient for orbital periods of at most 3~d.   We explain these results further in section~\ref{sec:convectionjupiters} by discussing the evolution of the convective timescale in the planet.   Finally, we summarize and discuss our results in section~\ref{sec:discussion}.

\section{Energy dissipation and circularization timescales} 
\label{sec:tcirc}

We consider a star of mass $M_c$ and a companion of mass $M_p$ in a binary system.   We note 
$\omega_{\rm orb}$ and $P_{\rm orb} = 2 \pi  / \omega_{\rm orb}$ the orbital frequency and period, respectively.    We start by reviewing results from paper~I, before calculating the circularization timescale. 

\subsection{Energy dissipation}
\label{sec:energydissip}

The flow under consideration, which is a superposition of convective motions and tidal oscillations, is assumed to be incompressible (this is satisfied for the oscillations  when considering gravity modes, as done in this paper, whereas convection itself is not in reality incompressible).   
It was shown from first principles  in paper~I that, when the tidal period $P$ is small compared to the convective timescale $t_{\rm conv}$,  the tides excited by the companion exchange energy with the convective flow {\em via} the Reynolds stress at a rate per unit mass given by:
\begin{equation}
D_R = \left< u'_i u'_j \right> \frac{\partial V_i}{\partial x_j},
\end{equation}
where ${\bf u}'$ and ${\bf V}$ are the tidal and convective velocities, respectively.  The brackets indicate an average over a time large compared to the tidal period but small compared to the convective turnover timescale, and the subscripts refer to Cartesian coordinates.  Here, the Reynolds stress $-\rho \left< u'_i u'_j \right>$ is given by the correlation between the components of the tidal velocity, not that of the convective velocity. 
This term was derived by writing  energy conservation equations  for both the mean convective flow and the tidal fluctuations.  In these equations, the Lagrangian derivative of the kinetic energy is written as the divergence of a flux of momentum, which represents the work done by internal stresses (pressure force, viscous and Reynolds stresses), plus the work done by external forces, plus a term which represents viscous dissipation of energy, plus or minus $D_R$.    In some situations, when writing  conservation of energy, it is difficult to unambiguously identify the terms which are part of the flux of momentum and those which represent dissipation or exchange of energy between the different components of the flow.  
However, as we show in appendix~\ref{appendix1},  there is no ambiguity in the present context.

\subsubsection{Transfer of energy from the tides to the convective flow}

In paper~I, we assumed
$D_R>0$ but  did not justify it, other than by saying that observations show that energy is transferred from tidal oscillations to the convective flow: this is the reason why binaries circularize and the orbits of the moons of giant planets evolve the way they do.   This requires the integral of $\rho D_R$ over the convective regions of the star or planet to be positive, where $\rho$ is the mass density.     We note that this term does not actually have to be positive at all times, and we now  argue that  integrating $D_R$ over a timescale larger than the convective turnover timescale yields an energy dissipation rate consistent with assuming $D_R>0$.   This is because energy fed to the convective flow is transported to the stellar surface by the enthalpy flux, whereas energy fed to the tides does not get dissipated.  
As $D_R$ is proportional to the convective velocity gradient, 
 it is very likely that its  sign fluctuates over time on a timescale which is on the order of the convective turnover timescale $t_{\rm conv}$, and we now examine how that affects the exchange of energy.      Let us assume that  $D_R<0$ in some part of the flow domain for a period of time $\tau \sim t_{\rm conv}$.  This results in  energy being transferred from the convective motions to the tides, which yields an increase of the tidal amplitude.  As the perturbing mass exerts a torque on the tidal oscillation, an increase of the amplitude yields a change of orbital energy.  
Therefore, the energy transferred from convection to the tides during the time $\tau $ is stored in the orbit.    
 When the gradient of the convective velocity subsequently changes sign,  $D_R$ becomes positive and  this energy is transferred back to the convective motions, together with the additional  energy which is put in the tides by the perturber during  the time $\tau $ over which $D_R >0$.  As the energy transferred to the convective flow is transported by the enthalpy flux to the stellar surface,   where it is ultimately converted into a radiative flux (e.g., \citealt{Miesch2005}),   on a timescale $t_{\rm conv}$, it is lost from the system and cannot be fed back into the tides when the gradient of the convective velocity changes sign again.  In other words,  the energy going from the convective flow to the tides is always given back to the convective flow, but the energy going from the tides to the convective flow is never returned to the tides and is  eventually radiated away.  Therefore,  over a time larger than  $t_{\rm conv}$, all the energy is irreversibly transferred   from the tides to the convective flow.   This can also be expressed by saying that, although the integral of $\rho D_R$ over the volume of the convective zone may not be positive at every instant, it is positive when an average over a time lager than $t_{\rm conv}$ is done.  Since the  orbital evolution timescales are much larger than $t_{\rm conv}$, they can then be calculated  by assuming that $D_R>0$ at all times.  

The argument above implies that tidal energy is irreversibly transferred from the tides to the convective flow if, after it has been extracted from the tides by convection,  it can be transported away efficiently by the convective enthalpy flux.  This requires convection to be important in the energy budget.   In the deep layers of the convective envelope of solar type stars, convection is inefficient and the radiative flux dominates.  Using MESA, we have calculated the radius $r_{\rm eq}$ at which the enthalpy flux becomes equal to the radiative flux (and therefore to about half the total flux, as these fluxes are the dominant contributions to energy transport).  For the Sun, a 10~Myr and a 9~Gyr solar mass stars, $r_{\rm eq}=0.75$, 0.6 and 0.75 stellar radius, respectively (in agreement with \citealt{Miesch2000} for the Sun and \citealt{Ballot2007} for the PMS star).  This is to be compared with the inner radius of the convective envelope, which is 0.73, 0.5 and 0.7 stellar radius, respectively, for these three stars.  Excluding the parts of the convective envelope below $r_{\rm eq}$ for the calculation of the tidal energy dissipation rate  increases the eccentricity damping timescale by a factor $\eta$ which increases with $P_{\rm orb}$ (as the regions where the tidal period is smaller than $t_{\rm conv}$, and which contribute most to dissipation, move towards smaller radii at longer $P_{\rm orb}$).  For the Sun, we find that $\eta \sim 1.25$ 
 for $P_{\rm orb}= 10$~d.  This is however of no consequence because,   as will be shown  below,  circularization is mostly achieved before and after the MS.  For a 10~Myr and a 9~Gyr solar mass stars,  $\eta \sim 1.02$ and 1.3, respectively,  for the largest orbital period of 17~d considered here.  
Given the uncertainties in the model, these differences in the eccentricity damping timescales are not significant.  Therefore, in this paper, we will calculate 
tidal energy dissipation over the whole convective envelope,   instead of just over the region above $r_{\rm eq}$.  

\citet{Barker2021}  have recently claimed that the term $D_R$ does not contribute to tidal dissipation.  We comment on this study in appendix~\ref{appendix2}, where we  note that they misidentify the correct term responsible for ener\-gy transfer  between tides and convection.   As a consequence, their calculations in the anelastic approximation do not prove that the $D_R$  formulation is invalidated as an energy--loss coupling between tides and convection.   If anything, the simulations show that $D_R>0$ in this approximation!   \citet{Barker2021} discount the effect of $D_R$ by noting that it is cancelled by another term, which they claim should be included.  But this is true {\em only} for the calculation  of the {\em mean flow}, not for the fluctuations, the quantity of interest here.   In the anelastic approximation,  the  point remains that
 $D_R$ is the {\em only} term through which convection can extract energy from the tides.  This is discussed in more details in appendix~\ref{appendix2}.


\subsubsection{Energy dissipation rate}

 As in paper~I, and assuming $D_R>0$ following the argument presented above, we  approximate $D_R$ as:
\begin{equation}
D_R= \left( \left| \left< u'_r u'_{\theta} \right> \right| + \left|  \left< u'^2_r  \right> \right|  \right) \frac{V}{H_c} + \left( \left| \left< u'^2_{\theta} \right> \right| +\left| \left< u'^2_{\varphi} \right> \right| \right) \frac{V}{r} ,
\label{eq:DRtide1}
\end{equation}
where $(r, \theta, \varphi )$ is a spherical polar coordinate system centered on the star in which the tides are calculated, and $H_c$ is the scale on which the convective velocity varies.  Thereafter, we will use the mixing length approximation $H_c=2H_p$, with $H_p$ being the pressure scale height.     The total rate of energy dissipation, ${\rm d}E / {\rm d}t$,  in the convective regions of the star is  obtained by multiplying $D_R$ by the mass density $\rho$ and  integrating over the convective regions.      

We introduce the following integral:
\begin{multline}
I_1 \left( \omega_{\rm orb}, m, n \right)= \int_{\left( t_{\rm conv} > P_{\rm orb}/{n} \right) } {\rm d} r \; \rho(r)  \times \\ \left\{ 
\left[ r \xi_r (r)  \frac{{\rm d}}{{\rm d}r} \left( r^2 \xi_r (r) \right) + \alpha_m r^2 \xi^2_r(r) \right] \frac{V(r)}{H_c(r)} 
\right.
\\
\left. 
+ \frac{ \beta_m + 5m^2}{18}    \left[ \frac{{\rm d}}{{\rm d}r} \left( r^2 \xi_r (r) \right) \right]^2 \frac{V(r)}{r}
\right\},
\label{eq:I1}
\end{multline}
where $m$ and $n$ are two integers, $\alpha_m =8$ and $\beta_m=4$ for $m=2$, $\alpha_m =12$ and $\beta_m=36$ for $m=0$ (only these two values of $m$ will be considered).  Here, 
$\xi_r(r)$ is the radial part of the radial component of the tidal displacement, for which we use the equilibrium approximation:
\begin{equation}
\xi_r(r)=r^2 \rho \left( \frac{{\rm d} p}{{\rm d} r} \right)^{-1},
\label{eq:xir}
\end{equation}
with $p$ being the pressure.  The amplitude of the tidal displacement is $f \xi_r$, with $f=-GM_p/(4 a^3)$, where $G$ is the gravitational constant and $a$ is the separation of the system.  The domain of integration covers the region where $t_{\rm conv}>P = P_{\rm orb}/n$, where $P $ is the period of the tidal oscillation excited by the relevant term in the Fourier series decomposition of the tidal potential (see paper~I for details).   In principle, when calculating the total energy dissipation, we should add the contribution arising from the regions where $t_{\rm conv}<P$.  However, as shown in paper~I, this is negligible for the orbital periods of interest here.  The equilibrium tide approximation is  actually not very good  in  regions where the  Brunt--V\"ais\"al\"a frequency   is not very large compared to the tidal frequency  or, equivalently, where $t_{\rm conv} >P$, which are the regions we are interested in here, and this yields to overestimating   tidal dissipation by a factor of a few for close binaries \citep{Terquem1998,  Barker2020}.  However, 
 given the level of uncertainties in the calculations presented in this paper, this approximation is sufficient.  

The rate of energy dissipation for a non--rotating star in a circular orbit can then be expressed as:
\begin{equation}
 \frac{{\rm d}E}{{\rm d}t}  = \frac{3 n^2}{40} \pi \left(  \frac{M_p}{M_c+M_p} \right)^2 \omega^6_{\rm orb} I_1 \left( \omega_{\rm orb}, 2, 2 \right) ,
\label{eq:dEdtc}
\end{equation}
with $n=2$. 

If the orbit has a non--zero eccentricity $e$, terms proportional to $e^2$ have to be added.  If in addition the star rotates synchronously with the orbit,  only the terms proportional to $e^2$ contribute to ${\rm d}E / {\rm d}t$, and they have to be modified to take the star's rotation into account.  

\subsection{Evolution timescales}
\label{sec:evoltime}

The energy which is dissipated leads to a decrease of the orbital energy, and therefore to a decrease of the binary separation.  The characteristic  orbital decay  timescale  is given by:
\begin{equation}
t_{\rm orb} \equiv - a \left( \frac{{\rm d} a }{{\rm d} t} \right) ^{-1} = 
 \frac{M_c M_p}{M_c + M_p} \frac{\omega^2_{\rm orb} a^2}{  2 \left( {\rm d}E / {\rm d}t \right)}.
 \label{eq:torb}
\end{equation}
If the star  is synchronized, $ {\rm d}E / {\rm d}t  \propto e^2$ and the timescale is very long for small eccentricities.  If the star is not synchronized,  the dominant contribution to the rate of energy dissipation comes from the  $m=n=2$ term  in the Fourier series decomposition of the tidal potential, so that
 $ {\rm d}E / {\rm d} t  $ is given by equation~(\ref{eq:dEdtc}).

Through tidal interaction, the angular velocity $\Omega$ of the star of mass $M_c$ increases if it rotates with a period longer than  the orbital period (or is non--rotating).
Assuming a circular orbit,   the spin~up (or
 synchronization) timescale is given by:
\begin{equation}
t_{\rm sp} \equiv  - \left(  \Omega - \omega_{\rm orb}   \right) \left( \frac{{\rm d} \Omega }{{\rm d} t} \right) ^{-1} \simeq 
\frac{I \omega^2_{\rm orb} }{     {\rm d}E / {\rm d}t  },
\label{eq:tsp}
\end{equation}
where $I$ is the moment of inertia of the star and we have used $\Omega \ll \omega_{\rm orb}$, as these are the values of $\Omega$ which contribute most  to $t_{\rm sp}$.

For the parameters of interest here, tidal interaction yields a decrease of the eccentricity $e$ \citep{Goldreich1966}.  The eccentricity damping timescale is defined as: 
\begin{equation}
t_e  = - e \left( \frac{{\rm d} e}{{\rm d}t}  \right)^{-1}.
\end{equation}
(This is the timescale which was denoted $t_{\rm circ}$ in paper~I).  
To calculate $t_e$, we need to expand the tidal potential to non--zero orders in $e$.  An expansion to first order  is sufficient, as higher order terms lead to short timescales and  a rapid decrease of $e$.  Most of the circularization process is therefore dominated by the stages where $e$  is small (\citealt{Hut1981}; \citealt{Leconte2010} and discussion in section~\ref{sec:compsun}).
If the star of mass $M_c$  is non--rotating, this timescale is given by:
\begin{multline}
 \left( t ^{\rm nr}_e \right)^{-1}=  \frac{3 \pi}{10} \frac{M_p}{M_c +M_p}  \frac{\omega^4_{\rm orb}}{M_c a^2}   
\left[ -\frac{1}{2} I_1 \left( \omega_{\rm orb}, 2, 2 \right) \right.
\\ \left.
- \frac{1}{16}  I_1 \left( \omega_{\rm orb}, 2, 1 \right) + \frac{147}{16}  I_1 \left( \omega_{\rm orb},2, 3 \right) + \frac{1}{4}  I_1 \left( \omega_{\rm orb}, 0, 1 \right)  \right],
\label{eq:tenr}
\end{multline}
where the superscript `nr' indicates that the calculation applies to a non--rotating body. 

\noindent If the star of mass $M_c$ rotates synchronously,   the  timescale becomes:
   \begin{multline}
 \left( t ^{\rm sync}_e \right)^{-1}=  \frac{3 \pi}{10} \frac{M_p}{M_c +M_p}  \frac{\omega^4_{\rm orb}}{M_c a^2}   \times \\
\left[   \frac{73}{8}  I_1 \left( \omega_{\rm orb},2, 1 \right) + \frac{1}{4}  I_1 \left( \omega_{\rm orb}, 0, 1 \right)  \right]
  ,
\label{eq:tesync}
\end{multline}
where the superscript `sync' indicates that the calculation applies to a synchronous body.   

\subsection{Circularization timescale}
\label{sec:circtime}

As pointed out above, if the eccentricity of the binary is initially large, it decreases relatively fast at first until it reaches a value $\sim 0.1$.    During this evolution, the orbital period decreases as well (this will be discussed in more details  in section~\ref{sec:compsun}).  When $e  \sim 0.1$, the orbital decay and eccentricity damping timescales can be calculated using the expressions given  in 
section~\ref{sec:evoltime}, valid for $e \ll 1$.  The ratio  $t_{\rm orb}/t_e$ is  then on the order of a few if the star is non--rotating (see also paper~I), but becomes larger than 10 for $e=0.1$ and increases as $e^{-2}$ when $e$ decreases if the star rotates synchronously (which is likely to be the case as $t_{\rm sp} \ll t_e$).  This indicates that, during most of the circularization process, when $e$ is at most on the order of 0.1, the eccentricity decreases at fixed orbital period.  

Therefore, we define the circularization timescale $t_{\rm circ}$ as the time it takes for the eccentricity of the binary system to decrease from some initial value $e_0 \sim 0.1$ to a final value $e_f \ll e_0$ for a given orbital period $P_{\rm orb}$.  This timescale is then given by the implicit equation \citep{Khaliullin2010}:
\begin{equation}
 \int_{e_0}^{e_f}  - \frac{1}{e}  \frac{{\rm d} e}{{\rm d}t}  {\rm d}t  =  \int_{t_0}^{t_{\rm circ}}   \frac{{\rm d}t }{t_e \left(t, P_{\rm orb} \right)}  ,
 \label{eq:defintcirc}
\end{equation}
where  $t_0$ is the time at which the binary system has an eccentricity $e_0$, and we have made it explicit that $t_e$ depends on $P_{\rm orb}$ and also on $t$ through the structure of the star.  
 Expressions~(\ref{eq:tenr}) and~(\ref{eq:tesync}) for $t_e$, which are valid for $e \ll 1$,  are used to calculate the integral on the right--hand side of equation~(\ref{eq:defintcirc}).   \citet{Meibom2005} use $e=0.01$ as the threshold for circularization, so we take $e_f=0.01$.  The left--hand side of equation~(\ref{eq:defintcirc}) is then equal to $\ln \left( e_0/ e_f \right) =2.3$ (because of the ln--dependence, this is not very sensitive on the choice of $e_0$).  

Below, we calculate the circularization timescales for late--type binaries and for systems with a star and a hot Jupiter.  For stellar binaries, the starting time $t_0$ will be determined by calibrating the results so that they match the observed circularization timescale for PMS binaries.    For binaries comprising a giant planet, $t_0$ will be taken as the time at which the planet starts to be close enough to the star that tidal interaction is significant.  


\section{Circularization of late--type binaries}
\label{sec:sun}

  In this section, we consider a binary which consists of  two identical stars with $M_c=M_p=1$~M$_{\sun}$. 
  
  \subsection{Circularization timescales for late--type binaries}
  \label{sec:suncirc}
  
  With both stars contributing to eccentricity damping, the timescales $t_e$ calculated in the previous section have to be divided by 2.  Using MESA,
we calculate the structure of a 1~M$_{\sun}$ star as a function of time from the PMS to an age of  12~Gyr.
 Models are output at times $t_k$, with the integer $k$ varying from 1 to 333.   We then compute:
\begin{equation}
I \left( k, P_{\rm orb} \right) =  \int_{t_0}^{t_k}   \frac{{\rm d}t }{0.5 t_e \left(t, P_{\rm orb} \right)}  \simeq \sum_{j=k_{\rm min}  }^{k} \frac{\Delta t_j}{0.5 t_e \left( t_j, P_{\rm orb} \right)} ,
\label{eq:sumtcirc}
\end{equation}
with $\Delta t_j=0.5 \left( t_{j+1} - t_{j-1} \right)$ and where $t_{k_{\rm min}}  \equiv t_0$.    
The 
 circularization timescale $t_{\rm circ}$ for the orbital period $P_{\rm orb}$   is then the time $t_k$ corresponding to the value of $k$ for which $\left| I \left( k, P_{\rm orb} \right) -2.3 \right|$ is minimum.     This can also be expressed by saying that 
 $t_{\rm circ}$  is the age that the binary system has reached when its eccentricity becomes equal to 0.01, and the corresponding  $P_{\rm orb}$ is what \citet{Meibom2005} refer to as the circularization period for that age.  
We note $P_{\rm PMS}$ the circularization period of PMS binaries, which have an  age $t_{\rm PMS}$.
We then choose $k_{\rm min}$ such that our results match this  circularization period.   In other words, 
we determine $k_{\rm min}$ such that, for $P_{\rm orb}= P_{\rm PMS}$, $I(k, P_{\rm orb})$ reaches the value of 2.3 for $t_k \simeq t_{\rm PMS}$.

\begin{figure*}
    \centering
   \includegraphics[width=1.5\columnwidth,angle=0]{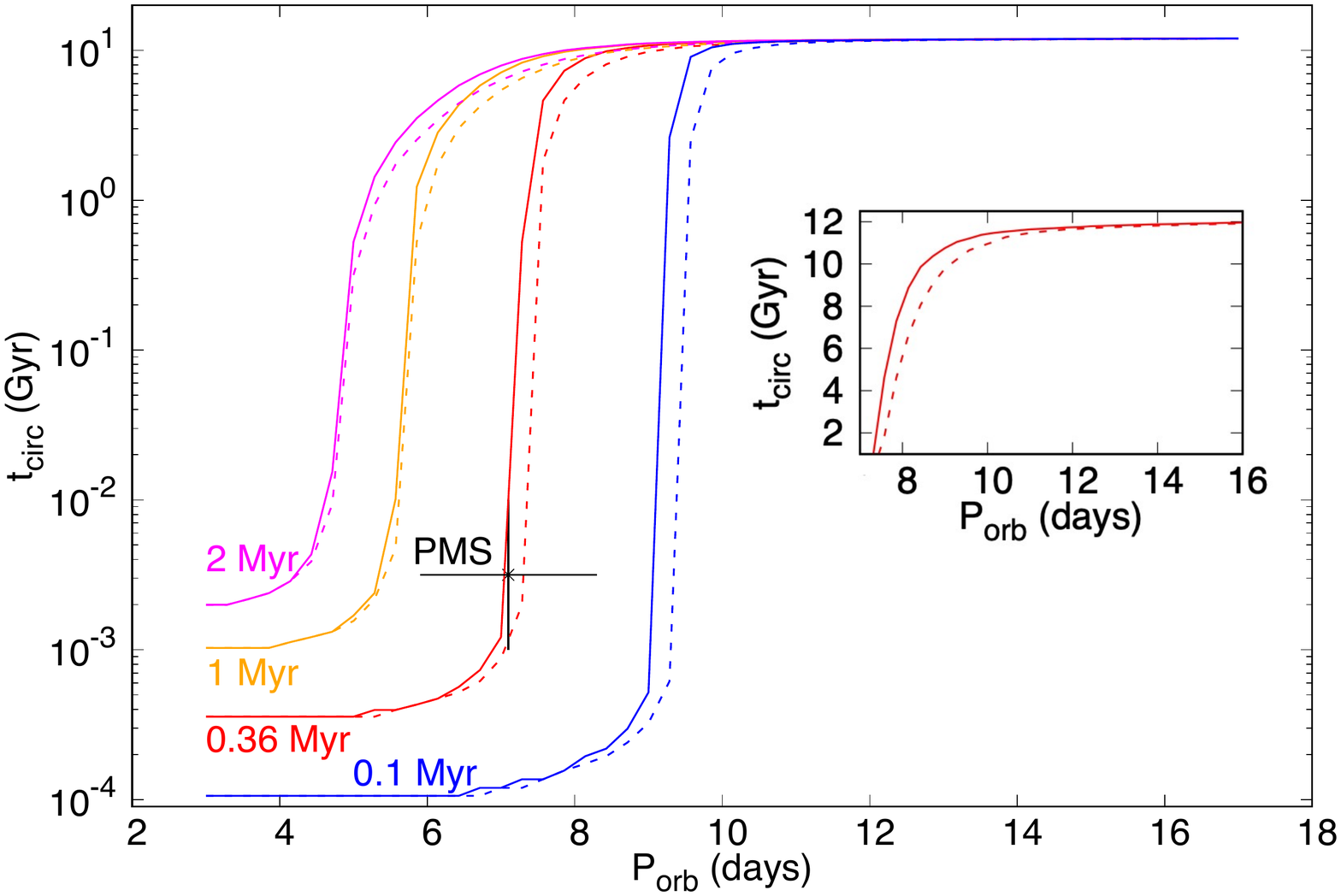}
    \caption{Circularization timescale for binaries with two identical solar mass stars.  Shown is $t_{\rm circ}$ in Gyr  for synchronized stars (solid lines) and non--rotating stars (dashed lines) and  using a logarithmic scale, {\em versus} orbital period in days.    
    The symbol with error bars represents the circularization period derived for PMS binaries by \citet{Meibom2005}.
  The starting time for the integration is $t_0=0.1, 0.36, 1$ and 2~Myr for the blue, red, orange and magenta curves, respectively.  The theoretical results match the circularization period for PMS binaries if  
  $t_0=0.36$~Myr.  The inset plot shows a zoom on $t_{\rm circ}$ between 2 and 12~Gyr for  the case $t_0=0.36$~Myr, using a linear scale.   
      }
    \label{fig7}
\end{figure*}

We calculate 
the
circularization timescale  for both non--rotating and synchronized stars,  which corresponds to replacing $t_e$ by $t^{\rm nr}_e$ or $t^{\rm sync}_e$, respectively, in equation~(\ref{eq:sumtcirc}).   These timescales are displayed in 
Fig.~\ref{fig7} for different values of $t_0$.  \citet{Meibom2005} have determined $P_{\rm PMS}=7.1$~d for a population of PMS binaries for which a representative age is $t_{\rm PMS}=3.6$~Myr, but with ages  spread between 1 and 10~Myr  \citep{Melo2001}.   We find that our results match this circularization period if we start the integration at   $t_0 \simeq 0.36$~Myr. 

As can be seen from Fig.~\ref{fig7}, tidal circularization is very efficient during the PMS phase, inefficient during the MS, and very efficient again on the RGB.  We  now discuss this in more detail. 

\subsection{Evolution of the convective zone}
\label{sec:convectionevol}

Equations~(\ref{eq:tenr}) and~(\ref{eq:tesync}) indicate that 
 the evolution of $t_e$  is determined by the time--dependence of $I_1$ (since $P_{\rm orb}$ is fixed, as discussed above).  Equation~(\ref{eq:I1}), in turn, shows that  $I_1$ depends on time through the evolution of the convective timescale $t_{\rm conv} \equiv H_c/V$,  the evolution of the extent of the convective zone,   and  that of the amplitude of the tidal displacement.    

\begin{figure*}
    \centering
   \includegraphics[width=1.5\columnwidth,angle=0]{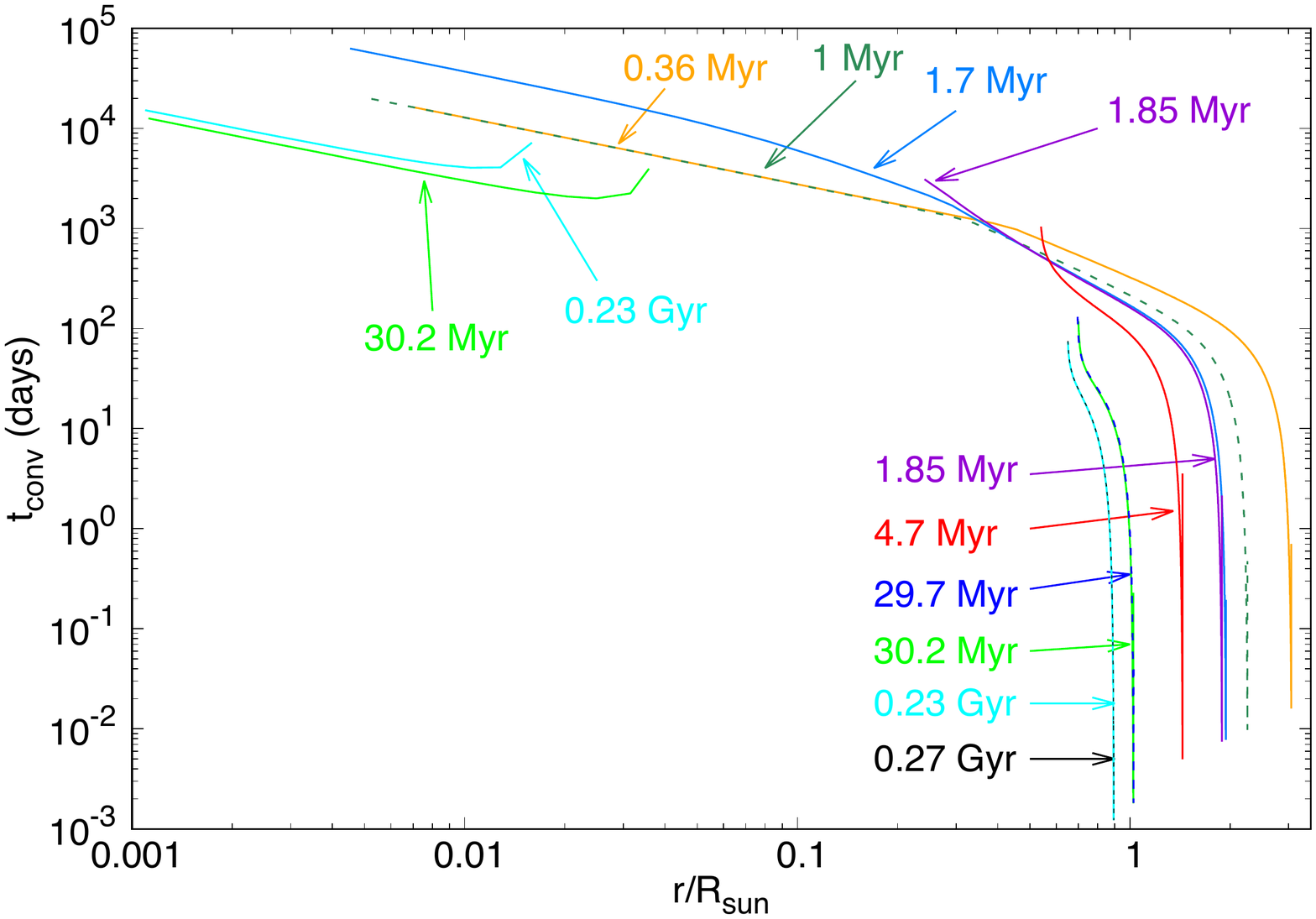}
    \includegraphics[width=1.5\columnwidth,angle=0]{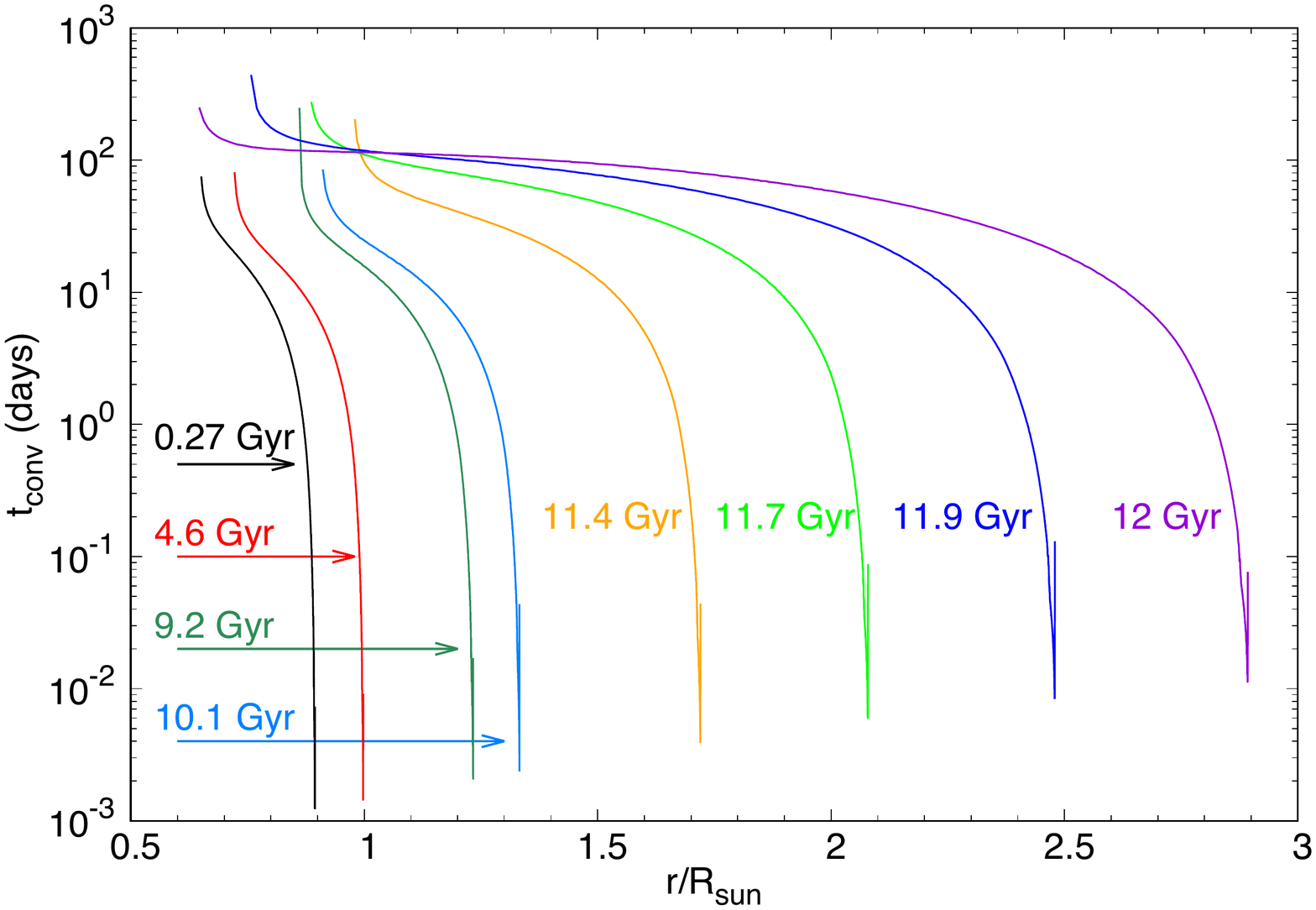}
    \caption{Convective timescale in a solar mass star.  Shown is $t_{\rm conv}$ (in days)  {\em versus} $r/{\rm R}_{\sun}$ in a 1~M$_{\sun}$ star for ages between 0.36~Myr and 0.27~Gyr (upper plot) and between 0.27~Gyr and 12~Gyr (bottom plot).  A logarithmic scale is used for $t_{\rm conv}$, and also for $r/{\rm R}_{\sun}$ in the upper plot.   Dashed lines are used to distinguish curves which are superimposed.
      }
    \label{fig3}
\end{figure*}

Fig.~\ref{fig3} shows  $t_{\rm conv}$ as a function of radius in a 1~M$_{\sun}$ star at different ages between 0.36~Myr  and 12~Gyr.   
 The youngest models have high opacities because of their low temperatures, so that  they remain fully convective as they contract down along  the Hayashi track   (the curves on the figure do not extend down into the very inner parts of the star because MESA does not output data for these regions).  By the time the star reaches about 1.8~Myr,  the temperature in the inner regions has increased sufficiently as a result of gravitational contraction that the opacities drop below the level where the stratification becomes stable.  Convection is then only present in an envelope that shrinks as the star evolves further \citep{Hayashi1966}. When the star reaches the zero--age main sequence (ZAMS), at $t \simeq 30$~Myr, the sudden output of energy  in the centre results in the inner parts becoming convective again.    This is due   to the large amount of energy released by the fusion of $^{3}$He into $^{4}$He as part of the pp--cycle.  
 This inner convective zone disappears  at $t \simeq 0.25$~Gyr, when $^{3}$He reaches its equilibrium abundance \citep{Chabrier1997}.     During the MS phase, the star is very stable and the temperature steady, so that the  structure evolves only moderately.   After the hydrogen in the core is exhausted, nuclear fusion in shells around  the core yields a swelling and cooling of the envelope, which is accompanied by an increase of the volume of the convective envelope as the star ascends the RGB.

\begin{figure*}
    \centering
   \includegraphics[width=1.5\columnwidth,angle=0]{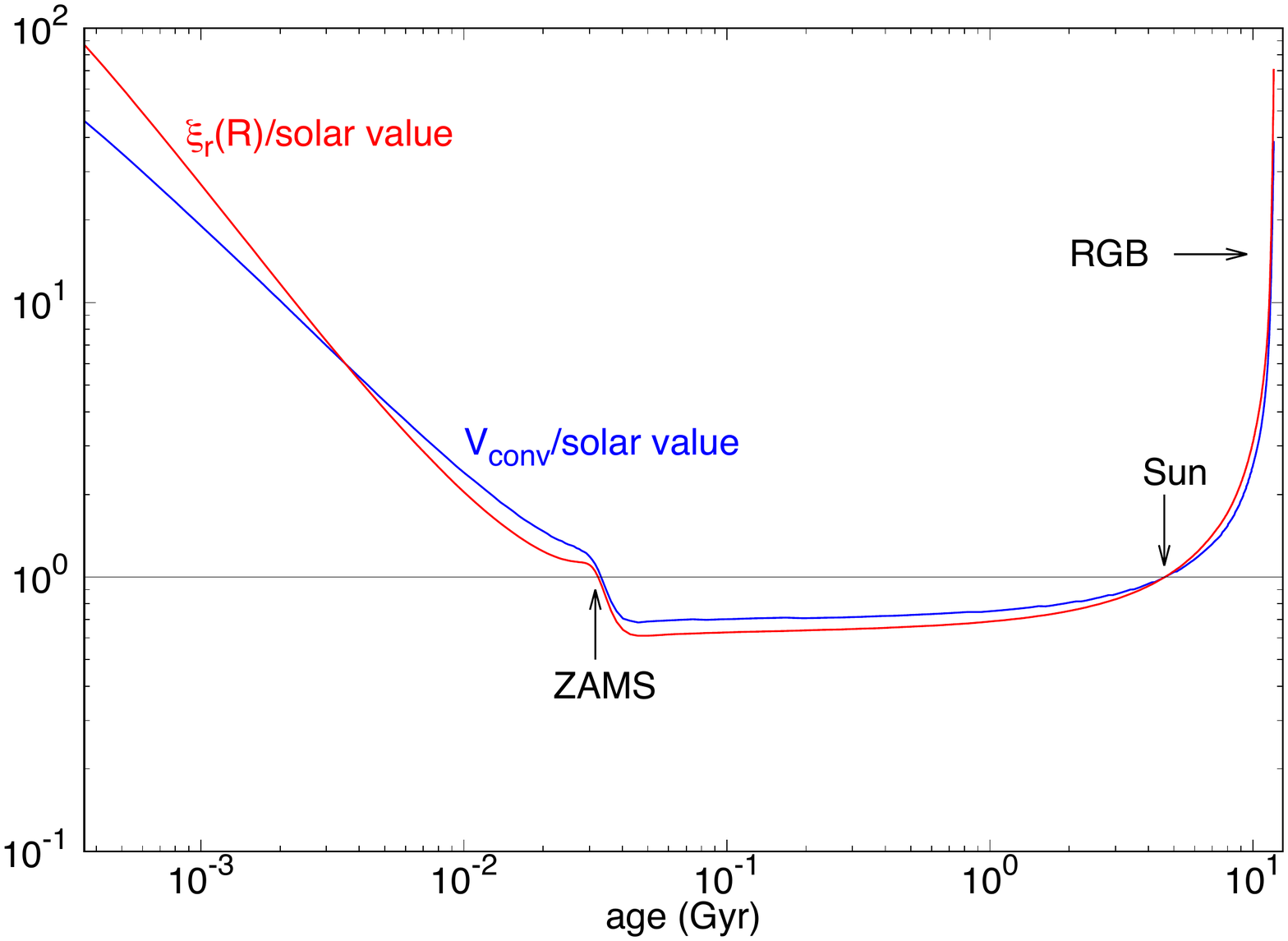}
    \caption{1~M$_{\sun}$ star models. Shown are $\xi_r(R)$ (red curve) and $V_{\rm conv}$ (blue curve) divided by the solar values {\em versus} age 
   in Gyr between $t_0=0.36$~Myr and 12~Gyr, using logarithmic scales.    The arrows indicate the location of the ZAMS, that of the actual Sun and the RGB. The horizontal line gives the values for the Sun.  
      }
    \label{fig2}
\end{figure*}

Convective regions located in the inner parts of the stellar interior do not contribute much to tidal dissipation, as the tidal displacement there is very small.  Therefore, rather than the mass, it is the volume of the convective zones which is a good indicator of the efficiency of tidal dissipation.  
Fig.~\ref{fig2}  shows the volume  $V_{\rm conv}$ of the convective regions and $\xi_r(R)$, where $R$ is the radius of the star, divided by the solar values, as a function of age.   We have also indicated on this figure  the location of the  ZAMS, at $t \simeq 30$~Myr, that of the actual Sun, at $t \simeq 4.6$~Myr, and the RGB, which starts at $t \simeq 10$~Gyr.
Using the equation of hydrostatic equilibrium,  equation~(\ref{eq:xir}) yields $\xi_r(r)=-r^4/ \left[ GM(r) \right]$, where $M(r)$ is the mass contained within the sphere of radius $r$.  Therefore, at $r=R$, and given that $M(R)=1$~M$_{\sun}$, $\xi_r(R)$ divided by the solar value is equal to $\left( R/{\rm R}_{\sun} \right)^4$.   At both $t=0.36$~Myr and $t=12$~Gyr, $R/{\rm R}_{\sun} \simeq 3$, so that the amplitude of the tide is about 80 times larger than in the Sun, for a given orbital period.    As can be seen from Fig.~\ref{fig2},  PMS stars have a convective region which is substantially larger than that of MS stars.  They also have a larger radius, which yields a larger tidal displacement  than in MS stars.   Tidal circularization is therefore very efficient during the PMS phase, but the efficiency decreases as the MS is approached.  During the MS, the volume of the convective regions does not change very significantly,  ranging from 0.7 shortly after the ZAMS to 2.6 at 10~Gyr, in units of the solar value.   The radius of the star also does not vary much, being between  0.9 and 1.3~${\rm R}_{\sun}$,  so that the amplitude of the tidal displacement stays roughly constant.  
In addition, as can be seen from Fig.~\ref{fig3}, the range of convective timescales in the star hardly varies on the MS.  Therefore, the eccentricity damping timescale $t_e$ is essentially independent of time on the MS.
 If we note $e_1$ and $e_2$ the values of the eccentricity at the   beginning and at the end of the MS, for a binary with orbital period $P_{\rm orb}$, we then have:
\begin{equation}
 \int_{e_1}^{e_2}  - \frac{1}{e}  \frac{{\rm d} e}{{\rm d}t}  {\rm d}t  =  \int_{t_1}^{t_2}   \frac{{\rm d}t }{t_e \left(t, P_{\rm orb} \right)}  \simeq \frac{t_2-t_1}{ t_e \left(t_{\rm sun}, P_{\rm orb} \right)},
 \label{eq:tMS}
\end{equation}
where $t_1 = 30$~Myr,  $t_2=10$~Gyr and $t_{\rm sun}$  is the age of the Sun (since $t_e$ is roughly constant, it can be approximated by the values calculated for the Sun).    For $P_{\rm orb}$ larger than 8~days, $t_e$ is larger than 10~Gyr, whether the stars are synchronized or non--rotating (see paper~I).  Therefore, $\left( t_2 - t_1 \right) / t_e < 1$, which implies that the eccentricity decreases only moderately, at best.   This means that  there is hardly any circularization happening during the MS.  By contrast, on the RGB, the volume of the convective zone increases very significantly, reaching 40 times the solar value at 12~Gyr.  The radius of the star also increases, leading to larger tidal displacements.  Tidal circularization is then efficient during that phase.  

\subsection{Comparison with observations }
\label{sec:compsun}

We now compare the circularization timescales calculated above with those derived from observations of binary populations of different ages.    The theory of tidal interactions predicts that stars synchronize on a timescale much shorter than the circularization timescale (see paper~I).  Other mechanisms outside the theory may prevent synchronization though,  as suggested by observations which show that a number of short period  eccentric binaries for which 
pseudo--synchronization would be expected on theoretical grounds  are   rotating either slower or faster   \citep{Lurie2017, Zimmerman2017}.   In any case,  as indicated in Fig.~\ref{fig7}, the circularization period we obtain is roughly the same for both non--rotating and synchronized stars, so that from here on we consider $t_{\rm circ}$ corresponding only to synchronized stars.     

\begin{figure*}
    \centering
   \includegraphics[width=1.5\columnwidth,angle=0]{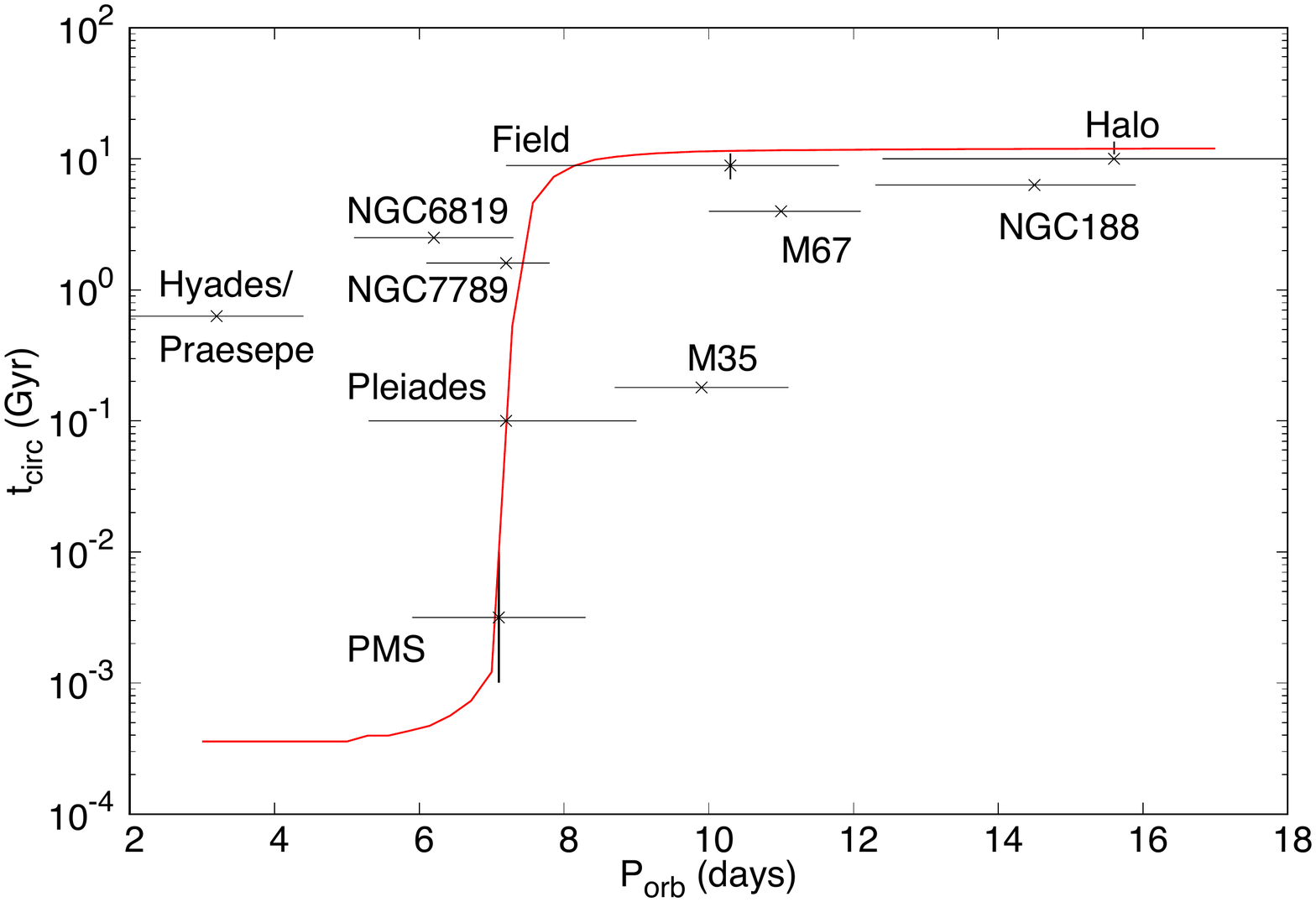}
    \caption{Circularization timescale for late--type binaries.  Shown is $t_{\rm circ}$  in Gyr  for synchronized stars, using a logarithmic scale, {\em versus} orbital period in days.    
    The symbols with error bars represent circularization periods derived from data using \citet{Zahn1977}'s circularization timescales \citep{Meibom2005}.
  The starting time for the integration, $t_0=0.36$~Myr, has been chosen so that $t_{\rm circ}$ matches the observed timescale for PMS binaries. 
      }
    \label{fig1}
\end{figure*}

Fig.~\ref{fig1} shows $t_{\rm circ}$ {\em versus} $P_{\rm orb}$  together with  observational data  for PMS and nine 
late--type   binary populations. 
For the Pleiades, Hyades/Praesepe, NGC188, the field and the halo binaries, the data   are from \citet{Meibom2005}.    
For M35, NGC6819, NGC7789 and M67, the data are from \citet{Leiner2015}, \citet{Milliman2014}, \citet{Nine2020} and \citet{Geller2021}, respectively.    The vertical bars for the field and halo binaries correspond to the spread in ages given by \citet{OMalley2017} and \citet{Duquennoy1991}, respectively.
 The starting time of the integration is chosen to be $t_0=0.36$~Myr, so that the theoretical result matches the observed circularization period of PMS binaries.  

The data for PMS, Pleiades, NGC7789 and 
NGC6819 binaries are consistent with the fact that circularization is efficient during the PMS but not on the MS.  The long circularization periods for the field and halo binaries are also consistent with circularization resuming towards the end of the MS, when the stars approach the RGB.    There are, however,  some notable discrepancies between the theoretical results and the observational data for the other clusters.  

To interpret these differences, it is important to recall how the circularization period $P_{\rm circ}$ is determined from the observed period--eccentricity distribution of the populations.  For all the populations represented in Fig.~\ref{fig1},  $P_{\rm circ}$ was obtained using the method  proposed by \citet{Meibom2005}, based on original arguments in 
\citet[see also \citealt{Mazeh1988}]{Duquennoy1992}.   The idea was 
that, since the binaries in a population of a given age had initially a Gaussian distribution of eccentricities,  significant eccentricities could still be found for relatively short period orbits after some time even when wider orbits had been circularized.   The argument was that binaries with initially high eccentricities would have only partially circularized.    This   has been used as a justification for not  identifying
 the circularization period with that of the shortest period eccentric orbit.    
 Moreover, since the longest period circular orbits would come from binaries with initially low eccentricities,  $P_{\rm circ}$ could not be identified with the period of those orbits either.  
\citet{Meibom2005} then proposed to define $P_{\rm circ}$ as being the orbital period at which a binary starting with the most frequent eccentricity of all the clusters, $e=0.35$,  reaches $e=0.01$, for a given age.
This circularization period is   determined  from the observations by fitting
  a function $e \left( P_{\rm orb} \right) $ to the period--eccentricity distribution.
The details of the function  are obtained by simulating a population of binaries and evolving their eccentricities using \citet{Zahn1977} eccentricity damping and orbital decay timescales (which are calibrated such as to fit the observations).   
 The error bars are related to the spread of $P_{\rm circ}$  obtained from different simulations for a given population.
 
A major limitation of this method is that it is based on a rate of eccentricity damping which is only valid to first order in $e$, and yields  timescales much too long when applied to  high eccentricities.      The argument of partial circularization for highly eccentric orbits is similarly based on this rate of eccentricity damping, and is actually incorrect, as we now discuss. 

\begin{figure*}
    \centering
   \includegraphics[width=1.5\columnwidth,angle=0]{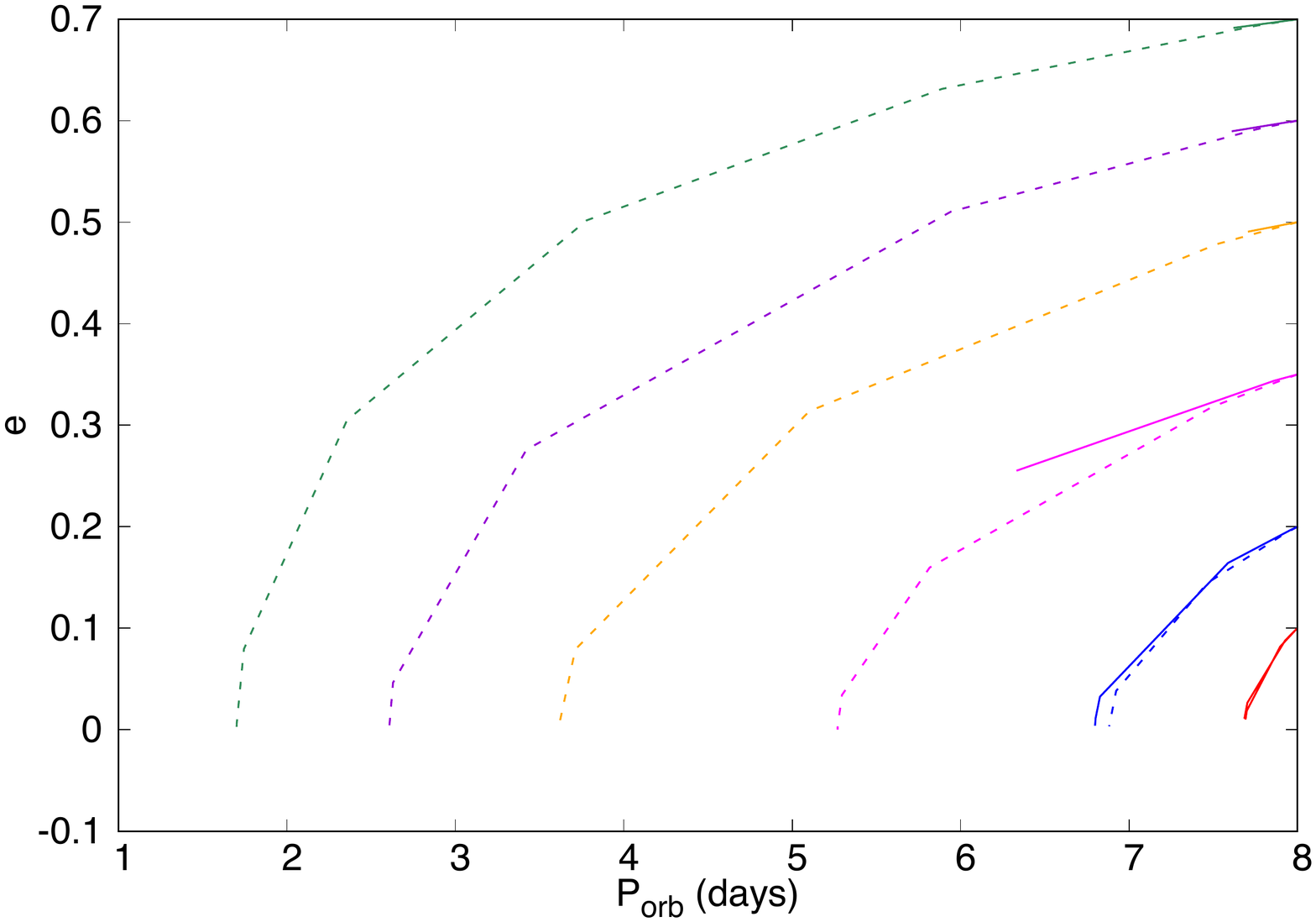}
    \caption{Comparison of \citet{Zahn1977} and \citet{Hut1981} timescales.  Shown is $e$ {\em versus} $P_{\rm orb}$, in days, for an initial period of 8~days using  the leading order equations~(\ref{eq:tez}) and~(\ref{eq:torbz}) from \citet{Zahn1977}  (solid lines) and equations~(\ref{eq:teh}) and~(\ref{eq:torbh}) from \citet{Hut1981} (dashed lines).  The different colors correspond to different initial eccentricities $e_0$ between 0.1 and 0.7.   The total time of integration is the same for all the curves, and is such that the eccentricity has decreased to 0.01 for $e_0=0.1$.  The small eccentricity approximation is only valid for $e \lesssim 0.3$, and leads to overestimating the orbital decay and eccentricity damping timescales by orders of magnitude if used for larger eccentricities.  
      }
    \label{fig8}
\end{figure*}

\citet{Zahn1977} theory for synchronized stars gives ${\rm d}e / {\rm d} t \propto e$ and ${\rm d} P_{\rm orb} / {\rm d} t \propto e^2$, which are the lowest orders in $e$ and therefore are restricted to  $e \ll 1$. 
  General expressions, valid to any order in $e$, were derived by \citet{Hut1981} in the context of the constant time lag model, and they can be written under the form:
\begin{multline}
\frac{{\rm d}e }{ {\rm d} t} =- \frac{18}{7}  \frac{1}{\tau_0} \left( \frac{P_{\rm orb}}{1 \; {\rm d}} \right)^{-16/3}
\frac{e}{\left( 1 - e^2 \right)^{13/2}}  \times \\ \left[  1+ \frac{15}{4} e^2 + \frac{15}{8} e^4 + \frac{5}{64} e^6 
\right. \\ \left.
- \frac{11}{18}  \left( 1 - e^2 \right)^{3/2}  \left(  1 + \frac{3}{2} e^2 + \frac{1}{8} e^4  \right)  \right],
\label{eq:teh}
\end{multline}
\begin{multline}
\frac{{\rm d} }{ {\rm d} t} \left( \frac{P_{\rm orb}}{1 \; {\rm d}} \right) =- \frac{6}{7}  \frac{1}{\tau_0} \left( \frac{P_{\rm orb}}{1 \; {\rm d}} \right)^{-13/3}
\frac{1}{\left( 1 - e^2 \right)^{15/2}}  \times \\ \left[  1+ \frac{31}{2} e^2 + \frac{255}{8} e^4 + \frac{185}{16} e^6  + \frac{25}{64} e^8 \right. \\ \left.
-   \left( 1 - e^2 \right)^{3/2}  \left(  1 + \frac{15}{2} e^2 + \frac{45}{8} e^4 + \frac{5}{16} e^6  \right)  \right],
\label{eq:torbh}
\end{multline}
where $\tau_0$ is the eccentricity damping timescale  at $P_{\rm orb}=1$~d and for small eccentricities, as defined below.   When $e \ll 1$, these expressions become, to lowest order in $e$:
\begin{align}
\frac{{\rm d}e }{ {\rm d} t} & =  -  \frac{e}{\tau_0} \left( \frac{P_{\rm orb}}{1 \; {\rm d}} \right)^{-16/3},  \label{eq:tez} \\
\frac{{\rm d} }{ {\rm d} t} \left( \frac{P_{\rm orb}}{1 \; {\rm d}} \right) & =  - \frac{57}{7}  \frac{e^2}{\tau_0} \left( \frac{P_{\rm orb}}{1 \; {\rm d}} \right)^{-13/3}, \label{eq:torbz}
\end{align}
which are the expressions given by \citet{Zahn1977} (with erratum in \citealt{Zahn1978}; note that \citealt{Meibom2005} have a factor of 3 instead of $57/7$ in eq.~[\ref{eq:torbz}]).  \citet{Duquennoy1992} calibrated these expressions by using $P_{\rm circ}=10.5$~d for the 5~Gyr old cluster M67.    Identifying the age with the eccentricity damping timescale, this yields $\tau_0=1.79 \times 10^{-5}$~Gyr.   Using equations~(\ref{eq:tez}) and~(\ref{eq:torbz}), these authors then calculated that a binary arriving on the MS with $P_{\rm orb}=11.6$~d and $e=0.54$ would reach $P_{\rm orb}=5.86$~d and $e=0.35$, which are the parameters for the binary KW181 in Praesepe,  after about 0.8~Gyr.   They concluded that the circularization period for this cluster could therefore be larger, as the eccentricity of KW181 was a result of incomplete circularization.  However, if we use the more accurate  equations~(\ref{eq:teh}) and~(\ref{eq:torbh}) instead, we find that,
after 0.8~Gyr,  the binary reaches $P_{\rm orb }=4.65$~d and $e=7 \times 10^{-6}$.    Using the leading order equations~(\ref{eq:tez}) and~(\ref{eq:torbz}) for $e \gtrsim 0.3$ leads to overestimating the  orbital decay and eccentricity damping timescales by at least an order of magnitude.  This is illustrated in Fig.~\ref{fig8}, which compares the evolution of $e$ as a function of  $P_{\rm orb}$ obtained when  using equations~(\ref{eq:tez}) and~(\ref{eq:torbz}) on the one hand, and equations~(\ref{eq:teh}) and~(\ref{eq:torbh}) on the other hand.

 Although \citet{Hut1981} equations~(\ref{eq:teh}) and~(\ref{eq:torbh}) have been derived for the constant time lag model, and therefore do not apply with the new formalism used in this paper, the fact that $e$ and $P_{\rm orb}$ decrease much more rapidly when  $e$ is large holds in general.  This can be understood by noting that at fixed orbital period,  the periapsis distance, which is proportional to $\left( 1-e \right)$, decreases when $e$ increases, and that most of the tidal interaction occurs when the stars are at periapsis \citep{Leconte2010}.  Therefore, if $e$ is large, it decreases very quickly at first and 
most of the circularization process is spent  circularizing orbits with $e \lesssim 0.1$, justifying our use of a constant $P_{\rm orb}$ in section~\ref{sec:circtime}.

The $e$--dependence of the eccentricity damping and orbital decay timescales   is a strong argument in favour of identifying $P_{\rm circ}$ with the period of the shortest period eccentric orbit, unless the probability of generating eccentricities in short period binaries through dynamical interactions with other stars is significant.  The latter was suggested as a possibility by \citet{Mazeh1990}, but ruled out as a general explanation by \citet{Meibom2005} as the presence of a third star is generally not detected.  In principle, such eccentricities could be produced by a star flying--by, rather than by a third member bound to the system,  in which case the perturber could have escaped and not be detectable.  However, as high eccentricities decrease rapidly through tidal interactions, the  fly--by would have  to have been a recent event if affecting short period orbits.
If $P_{\rm circ}$ were identified with the period of the shortest period eccentric orbit, we see from the data in \cite{Meibom2005} that the circularization periods of the PMS,  M35, M67 and NGC188 binaries could be smaller than the published values, while that of the field and halo binaries could be larger.  
 If $P_{\rm circ}$ for PMS binaries is indeed smaller than what has been suggested so far,  matching our theoretical results would simply require starting the integration at a  time $t_0$ longer than the value of 0.36~Myr considered above. 

 

\section{Hot Jupiters}
\label{sec:hotjupiters}

In paper~I, we considered binaries where the 
central mass was a solar type star and the companion a Jupiter mass planet.   As only the eccentricity damping timescale, and not the circularization  timescale, was calculated, 
  we revisit this problem in this section.  

 \subsection{Observations}
 \label{sec:jupitersobs}

Earlier papers by \citet{Halbwachs2005}, \citet{Pont2009} and \citet{Pont2011} indicated 
a circularization period of  5~d for binary systems comprising a MS solar--type star and a close--in giant planet.   More recent studies have confirmed that close--in giant planets tend to have circular orbits, with the data being consistent with a circularisation timescale of 1~Gyr for an orbital period of 3~d \citep{Bonomo2017}.

\subsection{Circularization timescale for hot Jupiters}
\label{sec:tcircjupiters}

We calculate the circularization timescale associated with the tides raised in the star by the planet in the same way as above, using $M_p=1$~M$_{\rm J}$.   

We also evaluate the circularization timescale associated with the tides raised in a Jupiter mass planet by the star, which corresponds to  $M_c=1$~M$_{\rm J}$ and $M_p=1$~M$_{\sun}$.     As long as the planet has not accreted all of its atmosphere, it has to maintain contact with the disc and is therefore unlikely to be very close to the star.   Evolution models by \citet{Marley2007} show that, when Jupiter finishes accreting all of its mass, which happens after $\sim 1$~Myr, it has a radius close to 1.4~R$_{\rm J}$, where  R$_{\rm J}$ is Jupiter's radius, and from that point onwards it can be reasonably well modelled as a contracting low--mass `star'.  In other words, the memory of the formation process (core--accretion model {\em versus} contraction of an object which starts with its full mass) is lost after about 1~Myr for a Jupiter mass object.   Therefore, we generate models of a 1~M$_{\rm J}$  planet using MESA, and shift the ages given by the code to re--assign the age of 1~Myr  to  the model which has a radius of 1.4~R$_{\rm J}$.  We  assume that the planet  gets close to the star immediately after it has finished accreting its mass, which corresponds to taking $t_0=1$~Myr in the calculation of $t_{\rm circ}$.  This gives  an upper limit on the amount of tidal dissipation.  We then calculate the circularization timescale as above,  evolving the planet with MESA for up to a few Gyr.  We also consider the case $t_0=2$~Myr for comparison. 

\begin{figure*}
    \centering
   \includegraphics[width=1.5\columnwidth,angle=0]{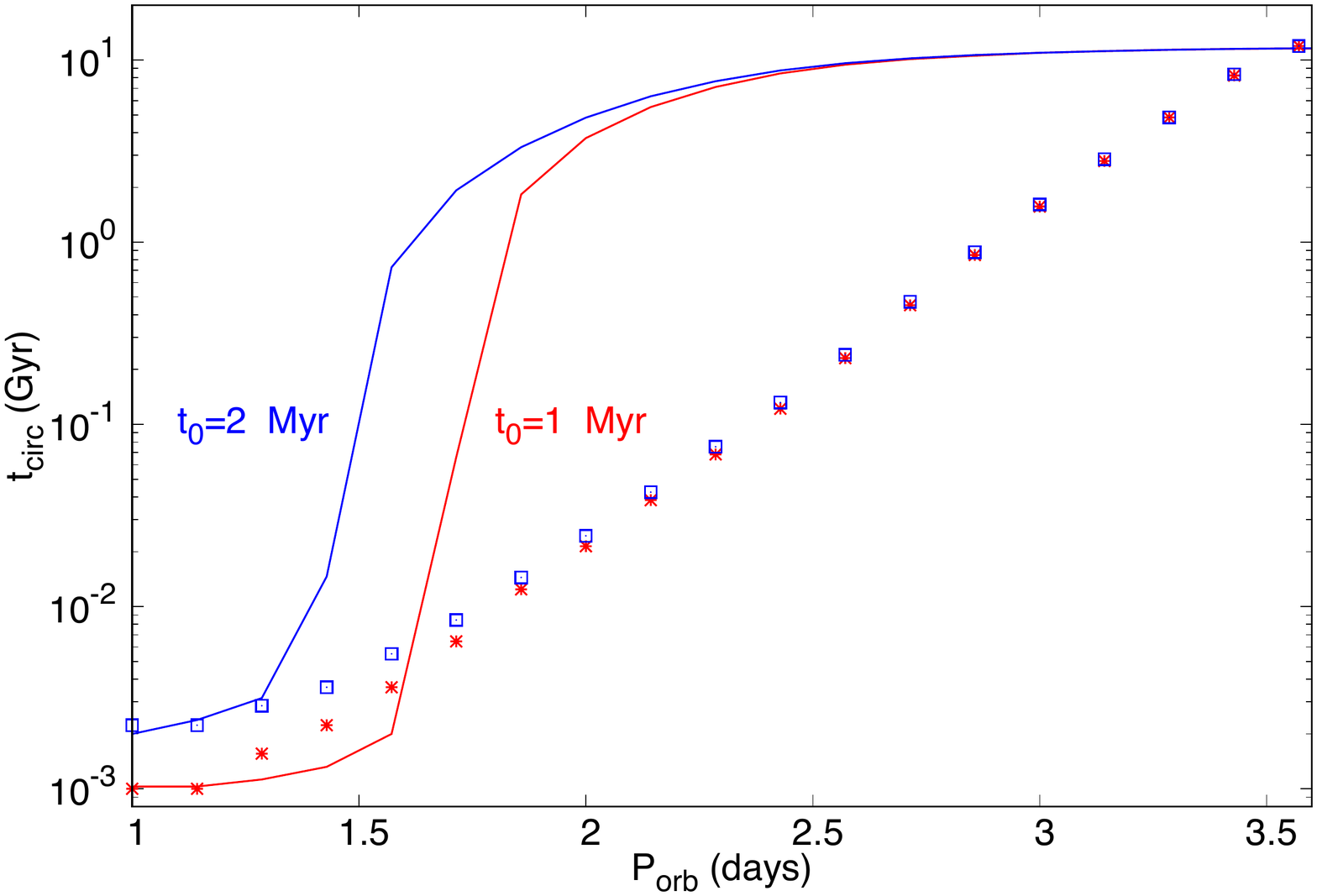}
    \caption{Circularization timescales for hot Jupiters.  Shown is $t_{\rm circ}$ (in Gyr)  associated with the tides raised in a 1~M$_{\sun}$ star  by a 1~M$_{\rm J}$  planet (solid lines) and  with the tides raised in a 1~M$_{\rm J}$ planet by a 1~M$_{\sun}$ star (symbols), using a logarithmic scale,  {\em versus} orbital period (in days).    The red and blue curves correspond to a starting time for the integration $t_0=1$ and 2~Myr, respectively.   For the parameters used here, the timescales are roughly the same whether the bodies are non--rotating or synchronized.  
    }
    \label{fig4}
\end{figure*}

These timescales are shown in Fig.~\ref{fig4} as a function of $P_{\rm orb}$.  For the parameters used here, $t^{\rm nr}_{\rm circ} \simeq t^{\rm sync}_{\rm circ}$, so we do not distinguish between non--rotating and synchronized objects.  Fig.~\ref{fig4} indicates that, if tidal interaction starts when the system  is 1~Myr old,  the total amount of energy dissipated in the star during the first few  Gyr can only circularize orbits with $P_{\rm orb} <2$~d.  Dissipation in the planet is much more efficient, circularizing orbits with $P_{\rm orb}$ up to 3.5~d in a few Gyr.

These results are in agreement with the observations described above.  

\subsection{Evolution of the convective timescale}
\label{sec:convectionjupiters}

As can be seen from Fig.~\ref{fig4},  there is ongoing circularization due to the tides in the planet and, to a lesser extent, to those in the star, during the MS.    

For the tides raised in the star, equation~(\ref{eq:tMS}) is still valid but, for the short orbital periods of interest here, $t_e<10$~Gyr, so that there is some eccentricity damping during the MS.  

\begin{figure*}
    \centering
   \includegraphics[width=1.5\columnwidth,angle=0]{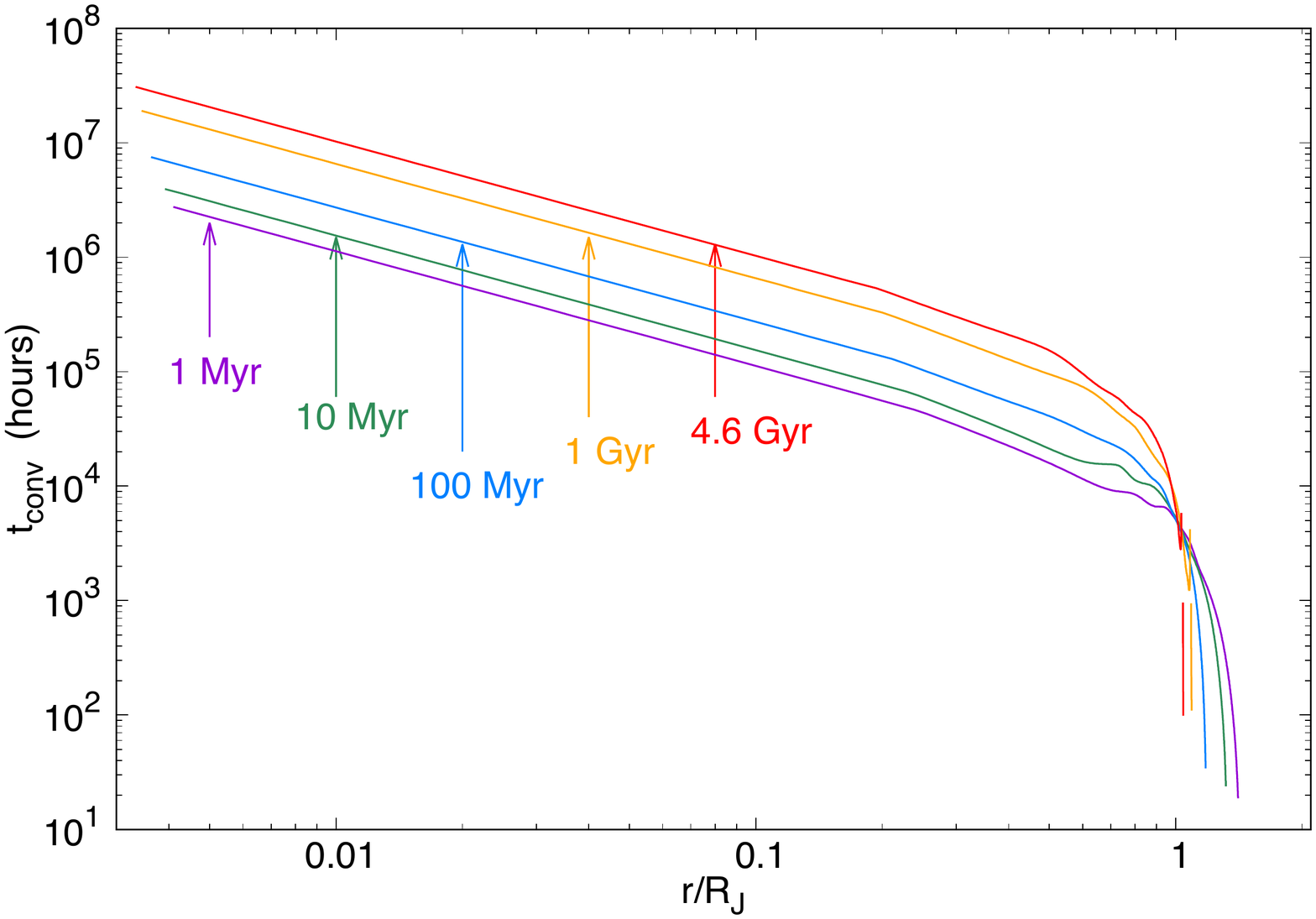}
    \caption{Convective timescale in a 1~M$_{\rm J}$  planet.  Shown is $t_{\rm conv}$ (in hours)  {\em versus} $r/{\rm R}_{\rm J}$, using logarithmic scales, at different ages.  From top to bottom, the different curves correspond to an age of 4.6~Gyr, 1~Gyr, 100~Myr, 10~Myr and 1~Myr, respectively.   
      }
    \label{fig6}
\end{figure*}

For the tides raised in the planet,  $t_e$ evolves with time due to  an increase of $t_{\rm conv}$.  This is  shown in Fig.~\ref{fig6}, which displays $t_{\rm conv}$ as a function of radius in a 1~M$_{\rm J}$  planet at different ages.   Between 1~Myr and 1~Gyr,  the power law $t_{\rm conv}  \propto t^{0.3}$ gives a crude fit to the results displayed in Fig.~\ref{fig6}.  As  $t_e \propto t_{\rm conv}$, this yields  $t_e \propto t^{0.3}$.   At $t=4.6$~Gyr, $t_e \propto P_{\rm orb}^{n}$ with $n \simeq 6$, as shown in paper~I  (the index of the power law was calculated for the tides raised in the star, but 
the tides raised in the planet yield roughly the same dependence 
 at short orbital periods).   Equation~(\ref{eq:defintcirc}) then yields $t_{\rm circ} \propto P^{8.6}_{\rm orb}$, which gives a crude fit to the results displayed in Fig.~\ref{fig4} for the timescales associated with the tides raised in the planet.

\section{Discussion and conclusion}
\label{sec:discussion}

We have calculated the circularization timescale of late--type binaries by integrating the inverse of the eccentricity damping timescale, $t^{-1}_e$, over time, starting at some time $t_0$.  The value $t_0=0.36$~Myr is required to  match the circularization period of  $7.1$~d which has been determined from the eccentricity-period distribution of  a population of PMS binaries  with ages between 1 and 10~Myr.  However, we have commented that the circularization periods determined from observations are very approximate, because they have been calculated using timescales valid only for small eccentricities.  Highly eccentric orbits circularize much faster than moderately eccentric orbits, so that the circularization period of a cluster is likely to be closer to that of the shortest period eccentric orbit than previously thought.   From \citet{Meibom2005}, we see that the shortest period eccentric orbit for the PMS population has a period of about 5~d.   To match this period  would require starting the integration at $t_0 \simeq 1$~Myr.   This time has to be interpreted as the time at which tidal interaction starts circularising the orbit.  Circularization may be prevented earlier on when the stars are still surrounded by a disc, as the interaction with a circumbinary disc may increase the eccentricity of the binary \citep{Artymowicz1991, Zrake2021}.   In this context,  the values of $t_0$ quoted above are consistent with the lifetime of discs being on the order of a few Myr.   

Our results show that tidal circularization is very efficient during the PMS phase, rather inefficient during the MS, and becomes efficient again when the stars approach the RGB.  To explain observations,  \citet{Mathieu1992} discussed what they called a hybrid scenario, in which circularization periods for populations with ages up to $\sim 1$~Gyr were obtained through tidal interaction during the PMS, while some more circularization was needed after that time to account for the longest circularization periods of older populations.  If interaction between tides and convection is the mechanism responsible for binary circularization, then, as we  have shown, and as had already been argued by \citet{ZhanB1989}, it cannot increase the circularization period during the MS.  This is because  the structure of the star, and therefore the eccentricity damping timescale $t_e$,  stays roughly constant during the MS.  As $t_e$ is larger than the age of the stars on the MS, there is hardly any circularization beyond what was achieved during the PMS.  However,  as the star reaches an age of 10~Gyr or so,  the convective envelope expands and circularization becomes efficient again.  

Our results are  in broad agreement with observations, to the extent that they match the circularization periods of the  PMS and Pleiades  binaries and that of the old populations such as the field and the halo.  Our results could still be adjusted to match those clusters if the circularization period of PMS binaries were reduced.  There is however a large scatter of circularization periods determined from observations of  clusters with ages between 0.2 and 6~Gyr, which is not consistent with our results.   Revisiting those circularization periods following the argument presented in this paper may lead to a better agreement.   

We have also calculated tidal dissipation in binaries containing a solar type star and a Jupiter mass  planet.  We have found that tidal interaction is dominated by dissipation in the planet, and yields a circularisation timescale of 1~Gyr for $P_{\rm orb}=3$~d, in agreement with observations \citep{Bonomo2017}.    A more detailed application of our formalism to hot Jupiters will be published separately.

 The calculations reported in this paper are based on the assumption that energy is transferred from the tides to the convective flow.  Although we cannot calculate $D_R$, as it involves the gradient of the convective velocity, we have argued that, over a timescale larger than the convective turnover timescale, all the energy is indeed lost from the tides.  This is because, even though energy may go back and forth between the tides and the convective flow, it is ultimately transported to the stellar surface by the enthalpy flux after it is returned to the convective flow.  Therefore, the orbital evolution timescales can be calculated by assuming $D_R >0$.   Such an interplay between tides  and convection could be simulated numerically by setting up a system which allows for  energy to be transported out of the  flow domain.  
 
 The results presented in this paper confirm that the new formalism introduced in \citet{Terquem2021} yields tidal energy dissipation rates which are consistent with the values expected for giant planets and late--type binaries,  perhaps solving a longstanding puzzle.  

\section*{Acknowledgements}
We thank Jonathan Fortney  and Robert Mathieu for kindly replying to some queries.  We are
  very grateful to Chris Mankovich for providing models of a young Jupiter and for pointing out that these models could be obtained using MESA.    We thank John Papaloizou for feedback on an early version of this paper. 
CT also thanks
 Steven Balbus for encouragements and very stimulating discussions.   Finally, we thank the referee for useful comments that have improved the manuscript. 
 This work used the {\em Modules for Experiments
in Stellar Astrophysics} (MESA) code available from {\em mesa.sourceforge.net}.

\section*{Data availability}

No new data were generated or analysed in support of this research.










\appendix

\section{Identifying the different terms in the energy conservation equation }
\label{appendix1}


Here, we show that the term $D_R$ can be interpreted unambiguously as the rate at which  energy per unit mass is exchanged between the mean flow and the fluctuations.    To this end, we retrace the steps that were outlined in paper~I in order to obtain the conservation energy equation for the mean flow.   Using the Reynolds decomposition in the $i$--component of Navier--Stokes equation, and averaging over a time long compared to the tidal period but small compared to the convective turnover timescales yields:
\begin{equation}
 \rho \frac{\partial V_i }{\partial
    t} +  \rho \left( {\bf V} \cdot \nab \right) 
      V_i 
  +  \rho \left< \left(   {\bf u}' \cdot \nab \right)  u'_i \right>= F_{{\rm p},i}
   + F_{{\rm visc},i}
+   \left< f_i \right> ,
   \label{eq:NSturb1}
\end{equation}
where ${\bf F}_{{\rm p}}$ is the average pressure force per unit volume,  $F_{{\rm visc},i}=  \partial S_{ij} / \partial x_j $ is the $i$--component of the average viscous force per unit volume, with $S_{ij}$ being the average viscous stress tensor, and ${\bf f}$ is the external force per unit volume acting on the flow (its average is zero if only the tidal force contributes).   
We have:
\begin{equation}
S_{ij} = \rho \nu \left( \frac{\partial V_i}{\partial x_j} +
  \frac{\partial V_j}{\partial x_i} \right).
\label{sigmaij2}
\end{equation}
Since the flow is incompressible, both the mean flow and the fluctuations are incompressible.  Assuming $\rho$ constant and interchanging the derivatives and averages, we then have:
\begin{equation}
\rho  \left< \left(   {\bf u}' \cdot \nab \right)  u'_i \right> \equiv  \rho \left< u'_j \frac{\partial u'_i  }{\partial x_j}      \right> 
=  \rho \frac{\partial  }{\partial x_j}   \left< u'_i  u'_j  \right> \equiv -  \frac{\partial R_{ij} }{\partial x_j} ,
\end{equation}
where $R_{ij}  \equiv -\rho \left< u'_i  u'_j  \right>$ is the Reynolds stress.   It is minus the average of the flux of the $i$--component of the fluctuating momentum transported along the $j$--direction by the fluctuations.  Therefore, $R_{ij}$ is the  $i$--component of the force exerted on a unit surface element which normals points in the $j$--direction  by the flow located in the region towards which the normal points.   By summing up the forces exerted  on the surface of a fluid element, we then obtain the next force per unit volume exerted on the fluid element as $F_{{\rm fluc},i} = \partial R_{ij} / \partial x_j$.  This force is due to the transport of fluctuating momentum by the fluctuations.    Equation~(\ref{eq:NSturb1})  can then be written under the form:
\begin{equation}
 \rho \frac{\partial V_i }{\partial
    t} +  \rho \left( {\bf V} \cdot \nab \right) 
      V_i  =
  F_{{\rm p},i}  + F_{{\rm visc},i} + F_{{\rm fluc},i}
+   \left< f_i \right> .
   \label{eq:NSturb2}
\end{equation}
Multiplying this equation by $V_i$, summing over $i$ and integra\-ting over some arbitrary  volume ${\cal V}$ of  fluid yields:
\begin{multline}
\iiint_{\cal V}     \frac{{\rm d} K}{{\rm d}t} {\rm d} \tau  =  
  \iiint_{\cal V}  V_i F_{{\rm p},i}  {\rm d} \tau   + W_{{\rm visc, vol}} + W_{{\rm fluc, vol}} 
+   \iiint_{\cal V}  V_i \left< f_i \right> {\rm d} \tau  ,
   \label{eq:NSturb3}
 \end{multline}
where:
\begin{equation}
\frac{{\rm d} K}{{\rm d}t} = \frac{\partial  K}{\partial
    t} +  V_j
  \frac{\partial K}{\partial x_j}  
\end{equation}
is  the Lagrangian derivative of the average kinetic energy per unit volume $K \equiv \rho V_iV_i/2$,  and we have defined:
\begin{equation}
W_{{\rm visc, vol}}  = \iiint_{\cal V}  V_i F_{{\rm visc},i} {\rm d} \tau , \; \; \; \; W_{{\rm fluc, vol}} = \iiint_{\cal V}  V_i F_{{\rm fluc},i} {\rm d} \tau    .
\end{equation}

\noindent Noting $\Pi$ the average pressure, and using the incompres\-sibi\-lity of the average flow, the first term on the right--hand side of equation~(\ref{eq:NSturb3}) can be written as:
\begin{equation}
  - \iiint_{\cal V}     V_i \frac{\partial  \Pi }{\partial x_i}   {\rm d} \tau 
  =  - \iiint_{\cal V}     \frac{\partial }{\partial x_i} \left(  \Pi  V_i \right)  {\rm d} \tau =  -\varoiint_{\cal S} \Pi {\bf V} \cdot {\rm d} \Sigma ,
   \label{eq:NSturb4}
\end{equation}
where ${\cal S}$ is the surface enclosing the volume ${\cal V}$.  This term represents the work done by the average pressure force on the surface of the volume of fluid.  Similarly, the last term on the right--hand side of equation~(\ref{eq:NSturb3}) represents the work done by the average external force $\left< {\bf f } \right>$ on the volume of fluid.  However, we now show that the other two terms do not  represent the work done by the viscous and Reynolds stresses.   
Indeed, the $i$--component of the viscous force exerted on a surface element ${\rm d} \Sigma $ is $S_{ij} n_j {\rm d} \Sigma$, where $n_j$ is the $j$--component of the unit vector normal to the surface.  Therefore, the work done by the viscous stress on the surface of the volume of fluid is:
\begin{equation}
W_{{\rm visc, surf}}  = \varoiint_{\cal S}  V_i S_{ij} n_j {\rm d} \Sigma =  \iiint_{\cal V}  \frac{\partial   }{\partial x_j}  \left(  V_i S_{ij} \right) {\rm d} \tau = W_{{\rm visc, vol}}  + D_{\rm visc} ,
\end{equation}
where we have defined:
\begin{equation}
D_{\rm visc} =  \iiint_{\cal V}  S_{ij}  \frac{\partial  V_i }{\partial x_j}   {\rm d} \tau .
\end{equation}
 In equation~(\ref{eq:NSturb3}),  $W_{{\rm visc, vol}} $
  can be interpreted as the work done by the net viscous force ${\bf F}_{\rm visc}$ acting on the {\em volume} of the fluid.  
For a small volume element moving with the bulk velocity ${\bf V}$, $W_{{\rm visc, vol}} $  is  equal to ${\bf V} \cdot {\bf F}_{\rm visc}$ times the volume, and this work results only in a change of the bulk velocity of the volume.  However, this is only part of the work $W_{{\rm visc, surf}}$ done by the viscous stress.  The other part, $D_{\rm visc}$, is related to the {\em deformation} of the volume element with no change of its bulk velocity.   This corresponds to energy which is irreversibly lost by  the mean flow: it is converted into thermal energy (this can be explicitly shown by writing an equation for the conservation of entropy).    The difference between $W_{{\rm visc, vol}} $ and $W_{{\rm visc, surf}}$  is due to the fact that the velocities vary across the volume element, and therefore the surface forces  are exerted at points which move with different velocities.   
Similarly:
 \begin{equation}
W_{{\rm fluc, surf}}  = \varoiint_{\cal S}  V_i R_{ij} n_j {\rm d} \Sigma =  \iiint_{\cal V}  \frac{\partial   }{\partial x_j}  \left(  V_i R_{ij} \right) {\rm d} \tau = W_{{\rm fluc, vol}}  + D_{\rm fluc} ,
\end{equation}
where we have defined:
\begin{equation}
D_{\rm fluc} =  \iiint_{\cal V}  R_{ij}  \frac{\partial  V_i }{\partial x_j}   {\rm d} \tau \equiv \iiint_{\cal V} - \rho D_R   {\rm d} \tau.
\end{equation}
The general form of the energy conservation equation has to be: rate of change of  kinetic energy equal flux of energy through the surface, which is associated with the work done by internal stresses at the surface,  plus work done by external forces plus $D$, where $D$ is the rate at which energy is irreversibly lost or gained by the flow.     This enables us to identify unambiguously $D_{\rm visc} $ as energy lost by the average flow (as it is positive definite) and 
 $D_{\rm fluc}$ as energy lost or gained depending on whether the term is positive or negative, respectively.   
 As this term enters the energy conservation equation for the fluctuations with the opposite sign, it represents the exchange of energy between the fluctuations and the convective motions {\em via} the Reynolds stress.  Energy is 
 transferred from the fluctuations to the convective motions if $D_{\rm fluc}<0$, that is to say if  the integral of $\rho D_R$ over the domain of the flow is positive. 

\section{Comments on Barker \& Astoul (2021) }
\label{appendix2}

 In a recent paper,  \citet[hereafter BA21]{Barker2021} claim to show that $D_R$ cannot contribute to tidal dissipation.    Their analysis is based on a study of Boussinesq and anelastic models.

We first note that BA21 misidentify the correct term responsible for the exchange of energy between the tide and convection.  The authors set  $I_{ee} \equiv \iiint \alpha {\rm d} \tau$ as the exchange term,  where $\alpha=\rho {\bf V} \cdot \left(  {\bf u}' \cdot \nab  {\bf u'} \right)$ and the integration is over the volume of the flow.   In the Boussinesq approximation,  and using BA21's notations, an integration by parts yields  $I_{ee}=-F+{ D}$, with ${ D}=\iiint \rho D_R {\rm d} \tau $ and $F= \iint  \left( \rho u'_i u'_j V_i \right) n_j {\rm d} \Sigma$, where the integrals are over the volume and the bounding surface of the flow, respectively, and $n_j$ is the j--component of the normal to the surface.   In the anelastic approximation, $I_{ee}=-F+{D}-t_1$, with  $t_1=  \iiint u'_i  \left( \partial \rho' / \partial t \right) V_i  {\rm d} \tau $ and where $\rho'$ is the density perturbation associated with the tide.  However, it can be shown that, although $F$ and $t_1$  contribute to the change of kinetic energy of the mean flow, they do not contribute to  that of the fluctuations.  Apart from the surface terms (which are zero in BA21 and therefore only redistribute energy within the flow), the only coupling between the tide and convection which changes the kinetic energy of the tide is the so--called deformation work ${D}$.  In other words, {\em the only term through which convection can extract energy from the tide is ${D_R}$}.

In the anelastic simulations,  BA21 find that ${ D} > 0$.  They go on to claim that there is no energy exchange between the tide and convection on the basis of $I_{ee}=-F+{D}-t_1=0$.    However, as we have just pointed out, $I_{ee}$ is not the correct term responsible for energy exchange between the tide and convection.  The fact that ${ D} > 0$ in this calculation would actually imply that kinetic energy {\em is} transferred from the tide to convection! 

In the Boussinesq approximation, BA21 obtain ${ D}=0$ by assuming that 
either 
(1)  
${\bf u'} \cdot {\bf n}={\bf V} \cdot {\bf n}=0$ along the bounding surface, where ${\bf n}$ is  the vector normal to the distorted surface, or
(2) 
${\bf V}=0$ everywhere on the bounding surface.

The condition ${\bf u'} \cdot {\bf n}=0$  is generally {\em incompatible with an object which radiates as a blackbody} (or with appropriate surface radiative boundary conditions).  When solving the stellar oscillation equations, the proper outer boundary conditions are i) that the surface is free, so that the Lagrangian variation of the pressure is zero, and  ii) the surface radiates as a blackbody,  which gives a relation between the Lagrangian changes in temperature and flux.  The point here is that the tidal displacement of the surface cannot be specified arbitrarily: it must adjust to ensure that these two surface conditions be satisfied, and this requirement in fact produces a nonzero component along the normal to the (distorted) surface.   It is {\it this} component which is responsible for the flux variation associated with stellar oscillations (Dziembowski 1977, Burkart et al. 2012, Bunting \& Terquem 2021).   
In principle,  a star could be mimicked by setting up a simulation with ${\bf u'} \cdot {\bf n}=0$,  if  some energy loss from the  surface were artificially added.  However, such a tidal displacement, which  does not satisfy the stellar oscillation equations, yields artificial constraints on global quantities, as demonstrated by the analysis of BA21, and is therefore not reliable.  

Assuming the impenetrability condition, ${\bf V} \cdot {\bf n}=0$, or the more restrictive no--slip condition, ${\bf V} =0$, 
 along  the bounding surface implies that there is no flux of tidal kinetic energy through the (upper or lower) surface of the convective zone.   Within the Boussinesq approximation, this means that no energy of any kind can be transported into or out of the convective zone.   The impenetrability or no--slip boundary conditions make physical sense  only if they are supplemented by an equation which enables the radiative heat flux to take over the transport of energy when the convective fluxes of enthalpy and kinetic energy diminish near the surfaces.   In fact, the sole emphasis on the kinetic energy flux is misplaced.   Kinetic energy transferred by the tide to the convective flow  becomes part of the overall kinetic energy of convection, and can be converted into thermal  energy via pressure acting on eddies which expand or contract.  This thermal energy  is then transported towards the surface of the Sun by the enthalpy flux  (Miesch 2005).  Then, as fluid elements move up, more and more of the thermal energy they contain is transported away by photons.   Therefore, even though there is no direct transport of kinetic energy through the surface of the convective zone, the tidal energy which is {\em initially} kinetic is transferred to the convective flow and escapes along with the rest of the energy that is already present in the convective zone.   This whole complex process, which controls the transport of energy in the Sun, can only be captured by solving the {\em full energy equation, including both kinetic and thermal energies}, not just the kinetic energy equation in isolation, as done by  BA21.
 
 The numerical simulations performed by BA21 using the Boussinesq approximation 
 and these artificial boundary conditions give the results predicted by their analysis.  In the simulations, ${\bf V}=0$  at the surface and the tide is restricted to be irrotational everywhere.  This set--up inevitably leads to the integral of $\rho D_R$ vanishing over the domain of the flow.
The simulations, by construction, can only confirm the unsurprising result that an irrotational tide cannot exchange energy with an incompressible flow enclosed within rigid boundaries.  This is, however, not relevant to stars or planets.   

Finally, we emphasise that the standard term $D^{\rm st}_R$, which couples the Reynolds stress associated with  the components of the convective velocity and the shear associated with the tide, {\em is not an alternative} to $D_R$.  When $P \ll  t_{\rm conv}$,  $\left< D^{\rm st}_R \right>=0$ and the only term through which convection may extract energy from the tide is $D_R$.

\color{black}

\bsp	
\label{lastpage}
\end{document}
